\documentclass[runningheads]{llncs}
\usepackage[T1]{fontenc}
\usepackage{color}
\usepackage{amsmath}
\usepackage{graphicx}
\usepackage{multirow}

\begin{document}
\title{Unsupervised Cardiac Segmentation Utilizing Synthesized Images from Anatomical Labels }
%
\titlerunning{Unsupervised Cardiac Segmentation}
%

\author{Sihan Wang\inst{1}, Fuping Wu\inst{1},  Lei Li\inst{2}, Zheyao Gao\inst{1}, Byung-Woo Hong\inst{3}, Xiahai Zhuang\inst{1}\thanks{Corresponding author (www.sdspeople.fudan.edu.cn/zhuangxiahai/)}} 
\authorrunning{Sihan Wang et al.}

\institute{School of Data Science, Fudan University, Shanghai, China \and
Institute of Biomedical Engineering, University of Oxford, Oxford, UK\\
\and
Computer Science Department, Chung-Ang University, Seoul, Korea\\
}
\maketitle              
\begin{abstract}
Cardiac segmentation is in great demand for clinical practice. Due to the enormous labor of manual delineation, unsupervised segmentation is desired. The ill-posed optimization problem of this task is inherently challenging, requiring well-designed constraints. In this work, we propose an unsupervised framework for multi-class segmentation with both intensity and shape constraints. Firstly, we extend a conventional non-convex energy function as an intensity constraint and implement it with U-Net. For shape constraint, synthetic images are generated from anatomical labels via image-to-image translation, as shape supervision for the segmentation network. Moreover, augmentation invariance is applied to facilitate the segmentation network to learn the latent features in terms of shape. We evaluated the proposed framework using the public datasets from MICCAI2019 MSCMR Challenge, and achieved promising results on cardiac MRIs with Dice scores of 0.5737, 0.7796, and 0.6287 in Myo, LV, and RV, respectively.
\end{abstract}
\section{Introduction}
Image segmentation is the core problem of medical image analysis, providing pixel-wise classification and precise location for further image analysis and clinic decision \cite{zhuang2020cardiac,zhuang2011framework}. Recently, great progress on semantic segmentation has been driven by the improvement of Convolution Neural Network (CNN) with the supervision of numerous annotated images \cite{yue2019cardiac,Unet:ronneberger2015u}. However, manual delineation is rather time-consuming and laborious. Hence, unsupervised segmentation algorithms are desired. 
There are plenty of efforts on unsupervised binary segmentation \cite{MS_CNN:kim2021unsupervised,melas2021finding}. 
However, extending these methods directly to multi-class situations could be difficult in both implementation and achieving satisfactory performance.
In this work, we study the multi-class unsupervised segmentation (MCUS) problem with both intensity and shape constraints. 

For MCUS, without supervision from annotated images, the design of loss function would be the  main determinant of segmentation performance. 
In the literature, many well-designed conventional energy functions have been proposed for pixel clustering \cite{MRF:panjwani1995markov,Cut:freedman2005interactive,Level:osher1988fronts}, 
such as the famous level-set based Mumford-Shah functional  \cite{MS:mumford1989optimal}, which can be implemented via neural network. 
Nevertheless, there are two inherent challenges among such efforts. 
Firstly, the number of segments, is an unknown parameter optimized by themselves \cite{MS:vese2002multiphase} and their non-convexity typically traps themselves in a local minimum, crippling the applicability to tackle multi-class segmentation \cite{MS:caselles1997geodesic,MS:massari1993finiteness}.
For example, Cai et al. \cite{MS_multi:cai2012image} utilized a clustering technique as post-processing on binary segmentation from the Mumford-Shah functional to generate multi-class results. To tackle this challenge, Vese et al. \cite{MS:vese2002multiphase} extended the energy function by simply representing multi-class results with multiply binary segments. In this work, we introduce a reasonable representing method utilizing the inherent relation among target objects to penalize the number of segments and implement it via U-Net.

Secondly, as mere intensity and length of regions are taken into account by these loss functions \cite{MS:caselles1997geodesic}, intensity-similar regions,  such as left and right ventricles, would be difficult to distinguish without any shape constraint. For example, Kim et al. \cite{MS_CNN:kim2019mumford} provided the shape constraint with extra annotated images in a semi-supervised training strategy.
Inspired by image-to-image translation in unsupervised domain adaptation \cite{tajbakhsh2020DA,chen2019unsupervised},  we propose the idea of providing synthesized annotated images for the segmentation network with a semi-supervised training strategy.  Specifically, we adopted the well-known Multimodal Unsupervised Image-to-image Translation (MUNIT) model \cite{MUNIT:huang2018multimodal} to predict synthetic images from given anatomical labels with adversarial loss compared with real MRIs.

However, the indelible gap between synthesized images and real images cripples the generalization ability of the segmentation network \cite{tajbakhsh2020DA}, leading to undesired predictions on real images. Hence, the network is desired to learn the latent content features under different appearance. Inspired by SimCLR \cite{chen2020simple}, we introduce an augmentation module to penalize the similarity of predictions from images in different perspective views as shape constraint.  

In this paper, we propose an unsupervised segmentation framework tackling the above-mentioned two challenges. The main contributions of our approach can be summarized as follows. 1) We extend Mumford-Shah functional for multi-class segmentation with relation constraint and implement it via neural network.
2) We propose a novel unsupervised segmentation training strategy, providing synthetic supervision from synthetic annotated images via an explicit image-to-image translation strategy. 3) We introduce intensity constraint and spatial constraint based on vanilla MUNIT for more realistic and precise predictions. 4) We introduce an augmentation invariant strategy to facilitate the model to generate the complex and multi distributions inherent in the shape of the data. and 5) We validate its performance with Cardiac MR images, and achieved promising results.

\section{Method}
\begin{figure}[!t]
    \centering
    \includegraphics[width=0.95\textwidth]{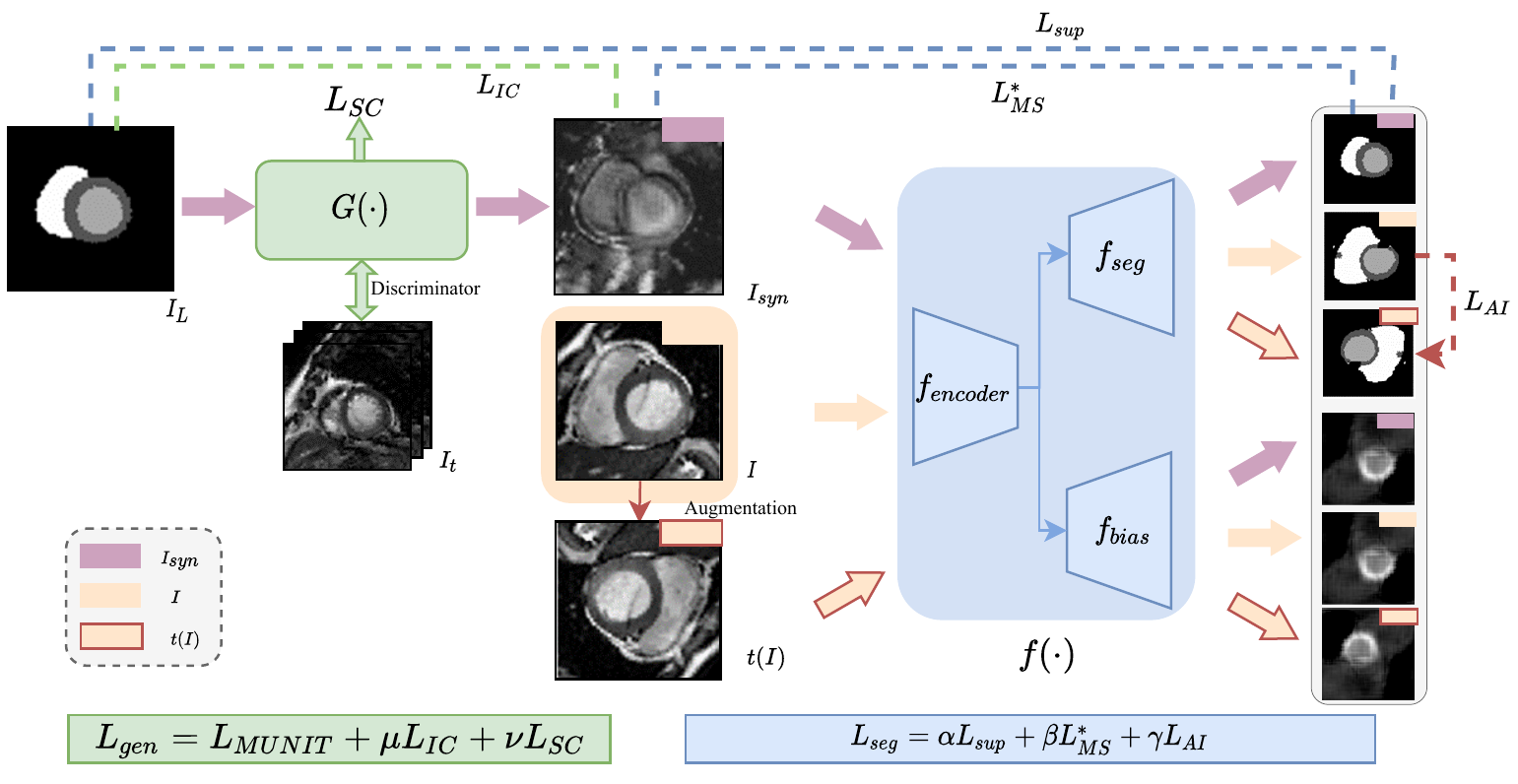}\\[-2.5ex]
    \caption{Overall structure of the proposed unsupervised segmentation framework.
    }
    \label{fig:my_label}
\end{figure}
In this work, we propose an unsupervised segmentation framework for multi-class delineation with both intensity and shape constraints. 
As illustrated in Fig.\ref{fig:my_label}, an intensity-oriented unsupervised loss function $L_{MS}^*$ is proposed based on Mumford-Shah functional \cite{MS:mumford1989optimal}, while synthetic annotated images $(I_{syn}, I_L)$ from generator $G(\cdot)$ provide shape supervision for segmentation network 
$f(\cdot)$. Moreover, self-learning module is also applied to emphasize structure.

\subsection{Unsupervised Segmentation Network with Intensity Constraint}
Let $I(x) \in R^{H \times W \times C}$ be an input image. 
We introduce an embedding function $\phi_n: \Omega \mapsto [0,1]$ for each class, and denote $\Phi = [\phi_1, \phi_2,...,\phi_N]$ as the encoding of prediction, where $N$ is the number of target classes. The unsupervised loss function $L_{un}$ for segmentation network contains three components, 
\begin{equation}
    L_{un} = L_{MS}^* + \eta L_{IR} + \varepsilon L_{smooth}, 
    \label{function_un}
\end{equation}
where $L_{MS}^*$ refers to variant Mumford-Shah functional, $L_{IR}$ is an inclusion regularization and $L_{smooth}$ is total variation loss as smooth regularization. Specifically, $L_{MS}^*$ is to penalize the intensity variance within a object,
\begin{equation}
    L_{MS}^* = \sum_{n=1}^{N} \int | I(x) - b(x)c_n |^2 \phi_n(x) dx,
\end{equation}
where 
$c_n$ is intensity representation of class $n$. A reduced form of it is the average intensity within the object. With regard of the heterogeneity of intensity within a class, a predicted bias field $b(x)$ is imposed on $I(x)$ to obtain a more precise intensity constraint and then $c_n$ is calculated as follows.
\begin{equation}
    c_n = \frac{\int I(x)b(x)\phi_n(x) dx }{ \int b(x)\phi_n(x) dx}.
    \label{C}
\end{equation}
The estimation of the bias field can be easily implemented in the segmentation network with extra convolution layers before the last segmentation layer (refer to $f_{bias}$ in Fig. \ref{fig:my_label}), which is activated by a sigmoid function. 
Moreover, 
 $L_{smooth}$ is to penalize the smoothness of generated bias field and the length of partitioning boundary \cite{MS:caselles1997geodesic}, and can be formulated as $L_{smooth} = \sum_{n=1}^{N} \int |\nabla \phi_n(x)| dx + \int |\nabla b(x)| dx$.
 
\begin{figure}[!t]
    \centering
    \includegraphics[width=0.8\textwidth]{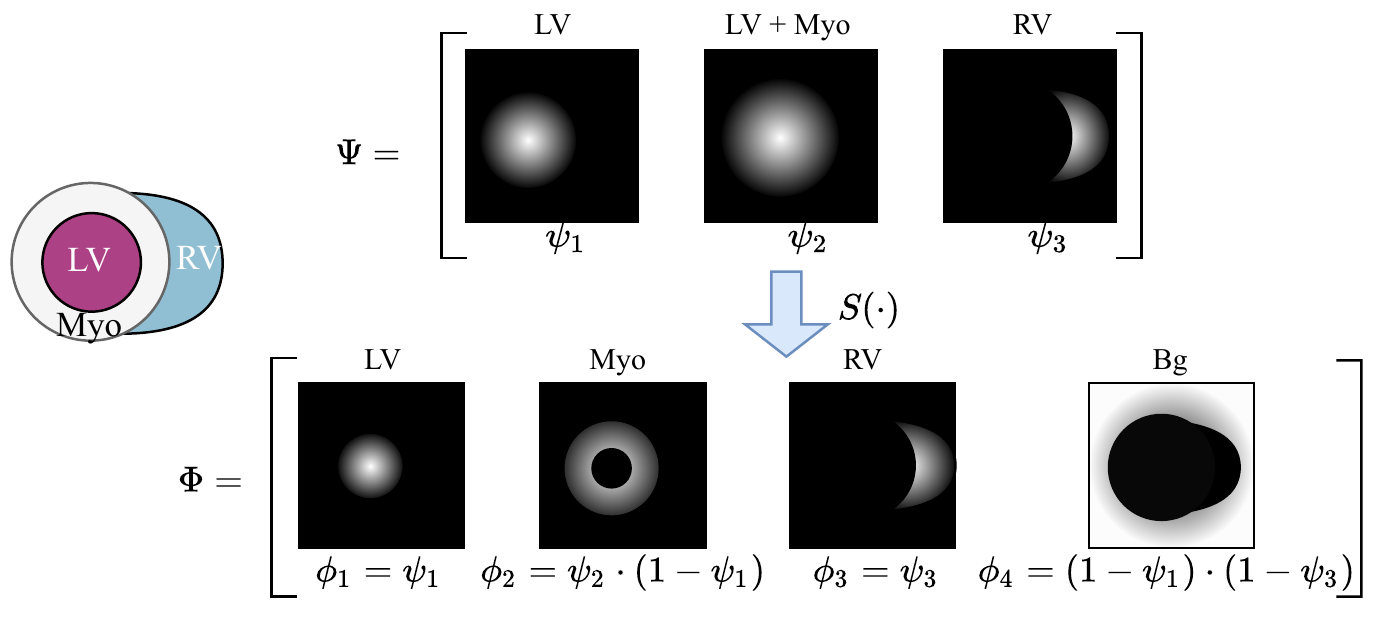}\\[-2.5ex]
    \caption{The illustration of mask mapping $S(\cdot)$ utilized in the Mumford-Shah function.}
    \label{fig:DataFormulation}
\end{figure}

As $L_{MS}^*$ and $L_{smooth}$ require the pre-definition of foreground subjects, for multi-class segmentation, we need to design an appropriate set of foregrounds, which should be compatible with each other, and easy to derive the final prediction via simple addition or multiplication.
To this end, we denote the output of the network as $\Psi$ with channel size  $N-1$, \textit{i.e.}, $\Psi = [\psi_1, \psi_2,...,\psi_{N-1}]$.
Specifically, for cardiac ventricle segmentation, as illustrated in Fig. \ref{fig:DataFormulation}, $\psi_1$, $\psi_2$ and $\psi_3$ are defined as LV, LV$+$Myo and Rv, respectively.
Then, with $\phi_1=\psi_1$ and $\phi_3=\psi_3$, the prediction for Myo and background can be respectively represented by $\phi_2 = \psi_2 \cdot (1 - \psi_1)$ and $\phi_4=(1 - \psi_2) \cdot (1 - \psi_3)$.
\setlength{\floatsep}{5pt plus 2pt minus 2pt}
\setlength{\textfloatsep}{5pt plus 2pt minus 2pt}
\setlength{\intextsep}{5pt plus 2pt minus 2pt}

 Furthermore, when solely  Mumford-Shah function is utilized, it is possible for $f(\cdot)$ to generate opposite $\psi_1$ and $\psi_2$. Hence, we introduce an inclusion regularization (IR) to penalize the inclusion relation between each foreground in $\Psi$,
\begin{equation}
    L_{IR} = || \psi_1 \cdot \psi_2 - \psi_1 ||^2_2,
\end{equation}
which constrains $\psi_1 \cap \psi_2$  to be $\psi_1$, and penalizes heavily when $\psi_1 \cap \psi_2 = \emptyset$.

\subsection{Strong Shape Constraints.}
We introduce the label-to-image translation technique and augmentation invariance for shape constrains.

\subsubsection{Label-to-image Translation} Let $i_l \in I_L$ be cardiac anatomic label map, and $i_t \in I_{t}$ be target CMR image. Note that the $I_L$ and $I_t$ are unpaired.
Our goal is to estimate $p(i_t|i_l)$ with image-to-image translation models $G(\cdot)$, which generates target image sample $I_{syn}$.
Our idea for shape constraint is that given $I_L$, the generated pair $(I_{syn}, I_{L})$ could help provide supervision for the segmentation model $f_{seg}(\cdot)$. For more precise prediction, we introduce intensity constraint (IC) and spatial constraint (SC) based on MUNIT \cite{MUNIT:huang2018multimodal}. 
Firstly, assume that pixels within an object should share similar intensity, then we can introduce an intensity constraint (IC) to emphasize such correlation as follows,
\begin{equation}
    L_{IC} = \int |I_{syn} - C|^2 \cdot I_L dx
\end{equation}
where C is calculated similar with Eq.(\ref{C}). 

Moreover, because of the significant variance in appearances and structures across slices of different positions in an image volume, we introduce a spatial constraint (SC) for the generator to force the outputs to be more discriminative. We categorize the slices into 5 classes according to their positions, denoted as $y$.
An additional fully-connection (FC) layers connected to vanilla content encoder $E^{c}$ \cite{MUNIT:huang2018multimodal} is then adopted to classify the position of the generated image,that is,
\begin{equation}
    L_{SC} = - \sum_{i}^{C} y_{i} \log(\hat{y}_{i})
\end{equation}
where $\hat{y}$ represents the position prediction.

The total loss for the generator is
\begin{equation}
    L_{gen} = L_{MUNIT} + \mu L_{IC} + \nu L_{SC}
    \label{L_gen}
\end{equation} 
where $L_{MUNIT}$ refers to the loss function of vanilla MUNIT, and $\mu$, $\nu$ are hyper-parameters.
By incorporating IC and SC, the generator can be penalized to be aware of the multi distribution and generate more realistic images in different positions.

\subsubsection{Augmentation Invariance.}The distance between the appearance of real image and $I_{syn}$ cripples the performance of the segmentation network. Hence, the augmentation invariant module is introduced to facilitate the network to sense the shape behind different appearances.

A set of stochastic data augmentation modules $\mathcal{T}$ is applied to randomly transform the input image, resulting in images in another perspective, \textit{i.e.},  $\tilde{I} = \mathcal{T}(I)$. $\mathcal{T}$ contains simple transformations, such as random rotation, random horizontal/vertical flip and random color distortions. Then the segmentation model $f(\cdot)$ is applied on the above pair $(I, \tilde{I})$. We assume that predictions of these homologue images remain the same, despite different orientations and contrasts. Augmentation invariant loss, $L_{AI}$ is utilized to encourage these predictions to be the same, 
\begin{equation}
    L_{AI} = \frac{1}{H\times W}(f(I) - t^{-1}(f(\tilde{I})))^2.
\end{equation}

\subsection{Overall Architecture}
As shown in Fig. \ref{fig:my_label}, there are two kinds of input images for segmentation network (blue part), i.e., target image $I(x)$ and synthetic annotated images $I_{syn}$, which is sampled from the generator (green part). The variant Mumford-Shah functional is applied as unsupervised supervision for segmentation. To utilize both synthetic annotated image $(I_{syn}, I_{L})$ and unlabeled data $I(x)$, the total loss function for the segmentation network is as follows:
\begin{equation}
    L_{seg} = \alpha L_{sup} + \beta L_{un} + \gamma L_{AI},
    \label{function_seg}
\end{equation}
where $L_{sup} = L_{DSC}(f_{seg}(I_{syn}), I_L)$ is the Dice loss for the prediction, $\alpha$ and $\beta$ are hyper-parameters with $\alpha = 0$ for unlabeled data. 
\section{Experiment}

\begin{table} [!t] \center
    \caption{The quantitative results of the ablation study and compared algorithm on CMR dataset. $\#1^*$ refers to $\#1$ with shape constraint via GAN. Syn refers to the network trained with synthetic cardiac images $I_{syn}$ and AI refers to the augmentation invariant module. Note that the Dice score of network with vanilla Mumford-Shah functional is not presented since
    only binary segmentation is achieved.}
	\label{tb:results}
			\resizebox{0.95\textwidth}{!}{ 
\begin{tabular}{c|ccc|ccc|c}
\hline
\multirow{2}{*}{Method} & \multicolumn{3}{c|}{Module} & \multicolumn{3}{c|}{Dice}                     & \multirow{2}{*}{\begin{tabular}[c]{@{}c@{}}Avg\\ Dice\end{tabular}} \\ \cline{2-7}
                        & $L_{MS}^*$    & Syn     & AI     & Myo           & LV            & RV            &                                                                     \\ \hline
\#1                     & $\surd$      & ×        & ×      & $0.2028\pm0.0991$ & $0.2914\pm0.1118$ & $0.1414\pm0.0188$ & $0.2119$                                                             \\
$\#1^*$                     & $\surd$      & ×        & ×      & $0.2110\pm0.0204$ & $0.3152\pm0.0859$ & $0.1746\pm0.0313$ & $0.2236$  \\
\#2                     & $\surd$      & $\surd$       & ×      & $0.5204\pm0.0606$ & $0.6799\pm0.1170$ & $0.3760\pm0.1330$ & $0.5254$                                                                  \\

\hline
MS\_CNN                 & -       &     -     & -      & $0.2430\pm0.0619$ & $0.122\pm0.1164$ & $0.2703\pm0.1165$ & $0.2118$                                                                 \\
\#3   & ×    & $\surd$         & $\surd$      & $0.5270\pm0.2218$ & $0.7049\pm0.22438$ & $0.1691\pm0.1758$ & $0.4670$   \\
Proposed                & $\surd$     & $\surd$        & $\surd$      & \textbf{0.5737$\pm$ 0.0870} & \textbf{0.7796$\pm$0.1075} & \textbf{0.6287$\pm$0.1320} & \textbf{0.6610} \\ \hline
\end{tabular}}
\end{table}

\subsection{Materials and Experimental Setups}
We evaluated the proposed algorithm on two datasets, i.e., synthetic images and public cardiac MR images. For synthetic datasets, we randomly generated images with square and circle in various size and position to evaluate robustness. For real cardiac image, we collected 45 bSSFP MRI subjects from the MICCAI2019 MSCMR challenge \cite{zhuang2020cardiac}, each of subject contains 8-12 slices.
We randomly divided these cases into 4 groups, i.e., 15 unlabeled training images $I(x)$, 10 anatomical labels $I_{L}$, 10 real target images $I_{t}$, and 10 testing images. The 2D slices extracted from original volumes were cropped into $128 \times 128$ and normalized via Z-score as the network inputs. Data augmentation including random resized crop, random rotation and flipping were applied.

The framework was implemented in Pytorch. 1) For the image translation network, we implemented the proposed module based on the official implementation of MUNIT \cite{MUNIT:huang2018multimodal}. The hyper-parameters $\mu, \nu$ in Eq. \eqref{L_gen} were both set to be 0 for the first 5k iterations, and 10 for the rest 15k iterations, while other hyper-parameters were kept the same as in \cite{MUNIT:huang2018multimodal}. The batch size for training was 1. 2) For the segmentation network, the network was pre-trained with synthetic images for 70 iterations with the batch size of 8, and then trained with both synthetic images and unlabeled images. The hyper-parameters $\beta, \gamma$ in Eq. \eqref{function_un} were 0.0001 and 0.001, respectively, while $\eta, \varepsilon$ in Eq. \eqref{function_seg} were both 1. The training was optimized with adam optimizer. The learning rate was set to $10^{-4}$. All experiments were implemented on one 24G NVIDIA TITAN RTX GPU.

\begin{figure}[!t]
    \centering
    \includegraphics[width=1\textwidth]{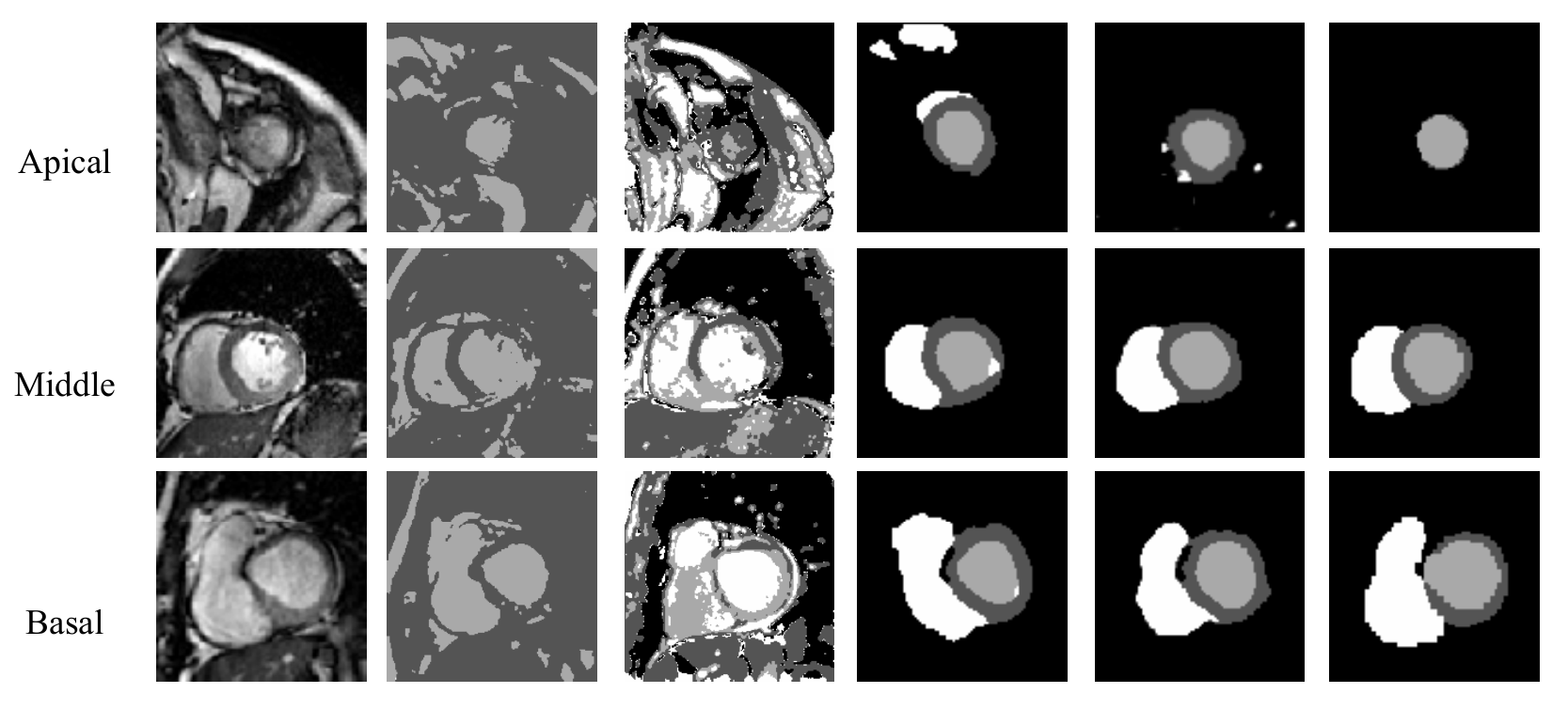}\\[-3ex]
    \caption{(a) Visualization of prediction on synthetic datasets. The third row represents the performance on image multiplied by bias field. (Since ground truths of synthetic images are obvious, we omit them.) (b) Examples from different spatial position are selected.}
    \label{fig:results_seg}
\end{figure}

\subsection{Performance of Segmentation Network}
We first provide a visual illustration of segmentation results on synthetic images to evaluate the performance of proposed $L_{MS}^*$. Space limited, the ground-truths of synthetic images are omitted due to their obviousness. The prediction of vanilla $L_{MS}$ and $L_{MS}^*$ are presented in the second and third rows in Fig. \ref{fig:results_seg} respectively. As one can see, merely binary predictions were generated with $L_{MS}$ while promising multi-class results could be achieved simply with $L_{MS}^*$. Moreover, it presents robustness of the segmentation network with images multiplied with bias field shown in the third row of Fig. \ref{fig:results_seg} (a). 

Table \ref{tb:results} and Fig. \ref{fig:results_seg} (b) present performance of the whole framework on CMR images. Besides the ablation study, we compared our framework with MS\_CNN \cite{MS_CNN:kim2019mumford}, which solved the non-convexity of Mumford-Shah functional theoretically and implemented it via U-Net. Our proposed algorithm achieved Dice scores of 0.5737,  0.7796, 0.6287 in Myo, LV and RV, respectively, which improved the performance by 40\% compared to MS\_CNN.
Since the intensity of RV is similar to that of LV, it is difficult for the network to distinguish them without strong shape constraint, which explains the unsatisfactory dice score in model \#1, \#2. With augmentation invariance (AI), there were improvements for the proposed network, especially in RV by almost 30\%. Moreover, as shown by \#3, without $L_{MS}^*$, the segmentor presents poor generalization and weak on distinguishing LV and RV.

As shown in Fig. \ref{fig:results_seg} (b), network with vanilla Mumford-Shah functional only predicted binary segments while multi-class segmentation could be predicted with the proposed $L_{MS}^*$. However, it was incapable of distinguishing outliers in similar intensity and had difficulty in distinguishing LV and RV. With the supervision of synthetic images, reasonable anatomic segmentation results could be achieved. Moreover, $L_{AI}$ dramatically improved the results, especially on RV. It is worth noting that the proposed framework could predict existing regions omitted in manual ground truth caused by expert experience or manual regulation, as shown in the first row of Fig. \ref{fig:results_seg} (b). 

\begin{table}[!t] \center
 \caption{The quantitative results of the ablation study and compared algorithm of generator. We evaluete the performance of the generator by Fréchet Inception Distance (FID). $\#1$ refers to basic MUNIT. Module IS and SC represent intensity constraint and patial
constraint present in section 2.2.}
	\label{tb:results_generator}
\begin{tabular}{l|ll|l}
\hline
\multirow{2}{*}{Method} & \multicolumn{2}{l|}{Module}  & \multirow{2}{*}{FID} \\ \cline{2-3}
                        & \multicolumn{1}{l|}{IC} & SC &                      \\ \hline
\#1                     & -  & -  & 355                  \\
\#2                     & $\surd$  & -  & 345                  \\
Proposed                & $\surd$ & $\surd$  & \textbf{320}                  \\ \hline
\end{tabular}
\end{table}

\begin{figure}[!t]
    \centering
    \includegraphics[width=0.8\textwidth]{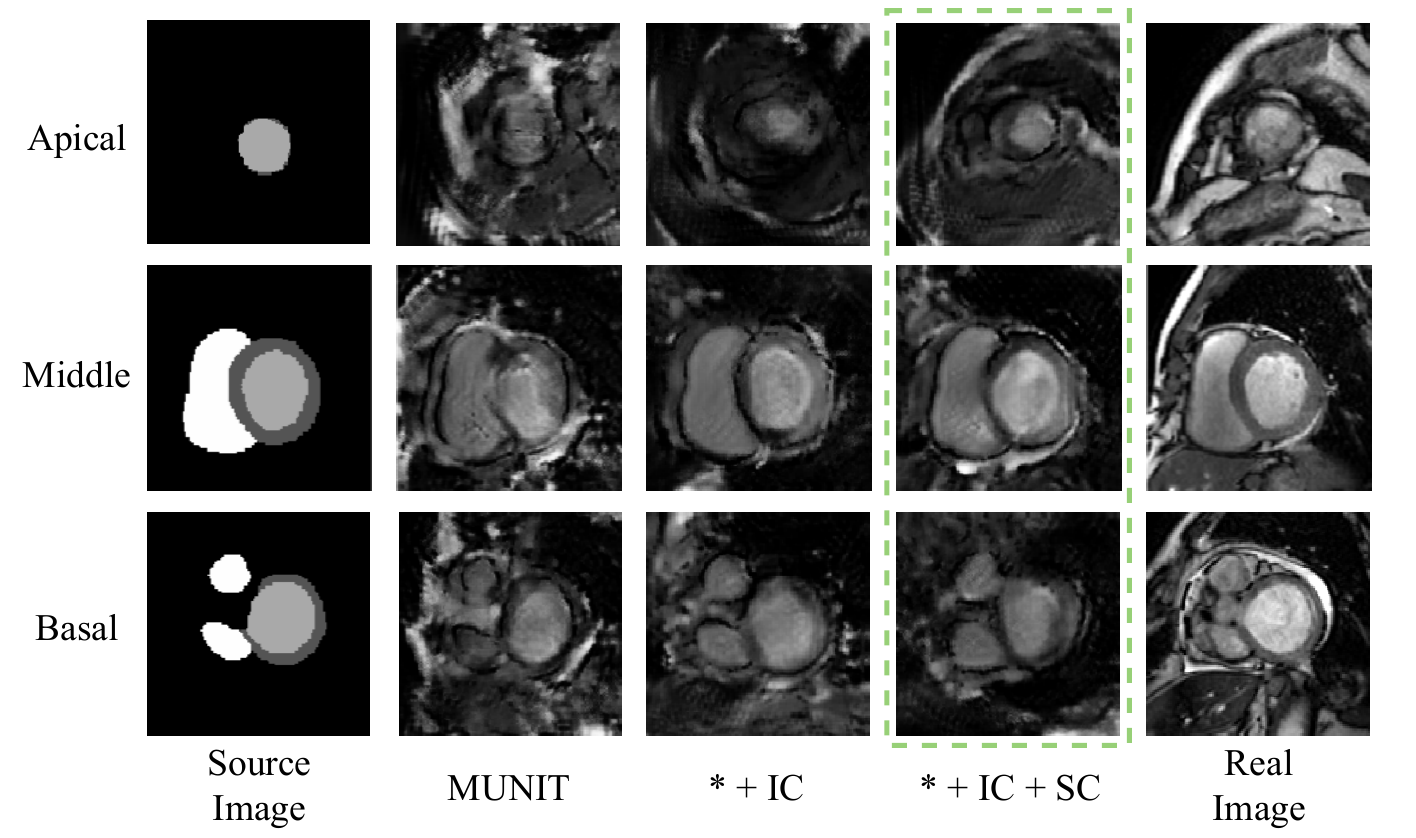}\\[-3ex]
    \caption{Visualization of the predictions from slices in different position. * refers to MUNIT while IC and SC refers to intensity constraint and spatial constraint module, respectively.}
    \label{fig:results_gen}
\end{figure}

\subsection{Performance of Generator}
We utilize the proposed Fréchet Inception Distance (FID) \cite{heusel2017gans} to quantitatively evaluate the performance of the generator. FID is to measure the similarity between two datasets of images and the lower its value, the more similarities there are between real and generated images. As shown in Table \ref{tb:results_generator}, the proposed generator with IC and SC constraint presents best performance. 
The visual performance of shape-to-image translation module is shown in Fig. \ref{fig:results_gen}. There were obvious drawbacks lying in the outputs of vanilla MUNIT, that is, the huge heterogeneity of intensity in the same region and the blurry boundary between myocardium and ventricle. Moreover, MUNIT presented a disability in generating apical and basal slices with clear structure. The application of IC module highly improved the intensity heterogeneity problem especially in the region of the myocardium. Additionally, spatial constraint dramatically improved the prediction in the apex.

\section{Conclusion}
In this work, we introduce a multi-class unsupervised segmentation framework with both intensity and shape constraints for cardiac anatomical segmentation. We firstly propose an intensity-based loss function capable for multi-class based on Mumford-Shah functional. In terms of shape constraint, two modules are introduced. Augmentation invariance penalizes the variance between predictions of the same image in different perspective views, and facilitates the segmentation network to learn the latent features preserving the information of shapes. A special and explicit label-to-image translation is proposed, generating synthetic images directly from given labels. This framework can provide strong shape constraint with synthetic supervision for the segmentation network. We evaluated the proposed framework on a synthetic dataset and real CMR images from a public dataset, and obtained promising Dice scores of 0.5737, 0.7796, and 0.6287in Myo, LV, and RV, respectively.

\section*{Acknowledgement}
This work was funded by the National Natural Science Foundation of China (No. 62011540404).
\bibliographystyle{splncs04}
\bibliography{reference}




\end{document}